\documentclass{sf2a-conf2012}
\usepackage{graphicx}
\usepackage{hyperref}
\usepackage[]{natbib}  
\usepackage[]{aeguill}
\usepackage{epstopdf}

\def\BibTeX{{\rm B\kern-.05em{\sc i\kern-.025em b}\kern-.08em
    T\kern-.1667em\lower.7ex\hbox{E}\kern-.125emX}}
\bibpunct{(}{)}{;}{a}{}{,}  

\begin{document}

\TitreGlobal{SF2A 2012}

\title{UV and optical polarization modeling of thermal\\active galactic nuclei : impact of the narrow line region}

\runningtitle{SF2A 2012}

\author{F. Marin$^*$}
\address{Observatoire Astronomique de Strasbourg, Universit\'e de Strasbourg, 
         CNRS, UMR 7550, 11 rue de l'Universit\'e, 67000 Strasbourg, France\\}
\thanks{$^*$ frederic.marin@astro.unistra.fr}

\author{R. W. Goosmann$^1$}

\setcounter{page}{237}

\index{Marin, F.}
\index{Goosmann, R. W.}


\maketitle

\begin{abstract}
In this research note, we start exploring the influence of the narrow line region (NLR) on the optical/UV 
continuum polarization of active galactic nuclei (AGN). We have upgraded our previous 3-component model of 
a thermal Seyfert nucleus that was composed of an equatorial, optically thin electron disc, a circumnuclear 
dusty torus, and a pair of collimated, optically thin electron winds \citep{Marin2012}. We have added a dusty 
extension with low optical depth to the outflows to account for continuum scattering and absorption inside 
the NLR. A spectropolarimetric comparison between our AGN models with and without NLR reprocessing is carried out. 
It turns out that the NLR can alter and even suppress the observed polarization dichotomy between type-1 and 
type-2 AGN. For type-2 AGN, it also significantly decreases the expected percentage of polarization and alters 
its spectral shape. While the NLR makes it more difficult to reproduce the observed polarization in type-1 
objects, it helps to explain spectropolarimetry observations of type-2 objects. Further studies including 
clumpy media need to be carried out to obtain more precise constraints on the polarization dichotomy and the 
reprocessing geometry of AGN.
\end{abstract}



\section{Introduction}
The unified model of active galactic nuclei (AGN) postulates that the emission of the continuum source and of 
the broad line region (BLR) is highly anisotropic because it is confined by the funnel of an obscuring dusty 
torus \citet{Antonucci1993}. A type-1 AGN has a visible BLR and is seen at a line of sight towards the central 
source that passes through the torus funnel. For a type-2 AGN, the view of the BLR is blocked by the torus body. 
The radiation from the center of the AGN can directly escape only along the polar regions of the funnel where 
it photo-ionizes conically shaped outflows. It was found that these winds are roughly stretched along the small 
scale radio-structure that is present in both radio-loud and radio-quiet AGN \citep{Wilson1983}. Beyond their 
sublimation radius, dust grains can form and coexist with the ionized outflow forming the so-called narrow line 
region (NLR). \citet{Capetti1999} showed that the NLR generally has a complex morphology consisting of filaments 
and compact emission knots (e.g. in NGC~4151) or narrow arcs (e.g. in Mrk~573). Due to this non-trivial geometry, 
it may be a difficult task to obtain accurate constraints on the NLR. However, spectropolarimetric studies of 
luminous AGN and numerical polarization modeling can improve our understanding if the complex radiative transfer 
between the various reprocessing regions is taken into account. 

\citet{Goosmann2007} investigated the continuum polarization induced by individual reprocessing regions, namely 
by dusty tori, polar outflows of various compositions, and equatorial scattering regions. Using this work as a 
starting point, we then modeled the optical/UV polarization emerging from a complex reprocessing system composed 
of three scattering components \citep{Marin2011,Marin2012}. The goal of the exhaustive study presented in \citet{Marin2012} 
was to put solid constraints on the conditions that reproduce the observed polarization dichotomy between type-1 
and type-2 AGN\footnote{Polarization is described as ``parallel'' when the $\vec E$-vector is aligned with the 
projected torus axis. We denote the difference between parallel and perpendicular polarization by the sign of the 
polarization percentage, $P$: a negative value of $P$ stands for parallel polarization, a positive $P$ for 
perpendicular one. Many type-1 AGN show parallel polarization while the polarization of type-2 objects is perpendicular
\citep{Antonucci1982,Antonucci1983}.}. So far, we have left out the NLR as it is expected to decrease the total 
amount of parallel polarization in type-1 AGN. For more details on this we refer the reader to \citet{Marin2012}. 
In this short paper, we start to investigate the effects on the continuum polarization that can emerge from adding 
to our previously model an NLR-like polar scattering region that is filled with optically thin dust.

\section{Model setup}
The continuum emitting region of our thermal AGN is modeled by an isotropic, point-like source of unpolarized 
radiation having a power-law intensity spectrum $F_{\rm *}~\propto~\nu^{-\alpha}$ with $\alpha = 1$. The source is 
closely surrounded by an electron-filled scattering ring producing the parallel polarization observed in type-1 
objects \citep{Antonucci1984}. At a larger distance and sharing the same equatorial plane, an obscuring, optically 
thick dusty torus blocks the radiation progressing towards edge-on directions. We assume that the torus collimates 
winds that are ejected from the inner accretion flow before they turn into a polar, hourglass-shaped outflow. 
These winds are optically thin, ionized and dominated by electron scattering. A complete description of the 3-component 
model can be found in Section~5 of \citet{Marin2012}. The model approximates the unified scheme of AGN and is 
optimized for the production of parallel polarization at a type-1 line of sight.

To study the effect of the NLR, we presently include extended, optically thin, dusty outflows sustaining the same 
half-opening angle and direction as the ionized winds. However, the amount of dust associated with the NLR is difficult 
to constrain from the observations due to 1) a complex filamentary structure between ionized gas and dust and 2) 
reddening corrections in the data reduction that can be a potential source of errors as the gas to dust composition 
is not known \citep{Cracco2011}. Taking into account that the fraction of dust in the NLR clouds must be smaller 
than in the circumnuclear region and in order to allow a large fraction of the radiation to escape in a polar direction, 
we fix the NLR opacity to a value significantly below unity \citep{Honig2012}. Table \ref{Table1} summarizes the 
parameters of the reprocessing regions in our models.

\begin{table}[]
  \centering
  {
   \footnotesize
   \begin{tabular}{|c|c|c|c|}
   \hline
      {\bf flared disk}               & {\bf dusty torus}                 & {\bf ionized outflows}               & {\bf NLR}\\
   \hline
      $R_{\rm min} = 3.10^{-4}$ pc    & $R_{\rm min} = 0.25$ pc           & $R_{\rm min} = 1$ pc               & $R_{\rm min} = 20$ pc\\
      $R_{\rm max} = 5.10^{-4}$ pc    & $R_{\rm max} = 100$ pc            & $R_{\rm max} = 10$ pc              & $R_{\rm max} = 70$ pc\\
      half-opening angle = 10$^\circ$ & half-opening angle = 30$^\circ$   & half-opening angle = 30$^\circ$    & half-opening angle = 30$^\circ$ \\
      equat. optical depth = 1        & equat. optical depth = 750        & vertical optical depth = 0.03      & vertical optical depth = 0.24\\
      electron scattering             & Mie scattering                    & electron scattering                & Mie scattering\\
   \hline
   \end{tabular}
  }
  \caption{Parameters of the individual scattering regions in our 3- and 4-component models. 
	   Note that for the polar outflow, the torus and the NLR, the half-opening angle 
	   is measured with respect to the vertical, symmetry axis of the model, while for 
	   the flared-disk the half-opening angle is taken with respect to the equatorial 
	   plane.}
  \label{Table1}
\end{table}

We apply the latest public version of {\it STOKES}, a Monte Carlo radiative transfer code including a treatment of 
the polarization. The code was initially presented in \citet{Goosmann2007} and new elements have been added by 
\citet{Marin2012}. The code is freely available on the Internet\footnote{http://www.stokes-program.info/}. 
It handles absorption and multiple scattering in a complex environment of emitting sources and reprocessing regions. 
Spectropolarimetric and polarization imaging results can be computed at any polar and azimuthal viewing angle. 
For more details on {\it STOKES} and its possible applications, we refer the reader to \citet{Marin2012} and 
references therein.

\section{Spectropolarimetric results}

\begin{figure}
   \centering
   \includegraphics[trim = 7mm 2mm 40mm 7mm, clip, width=15.8cm]{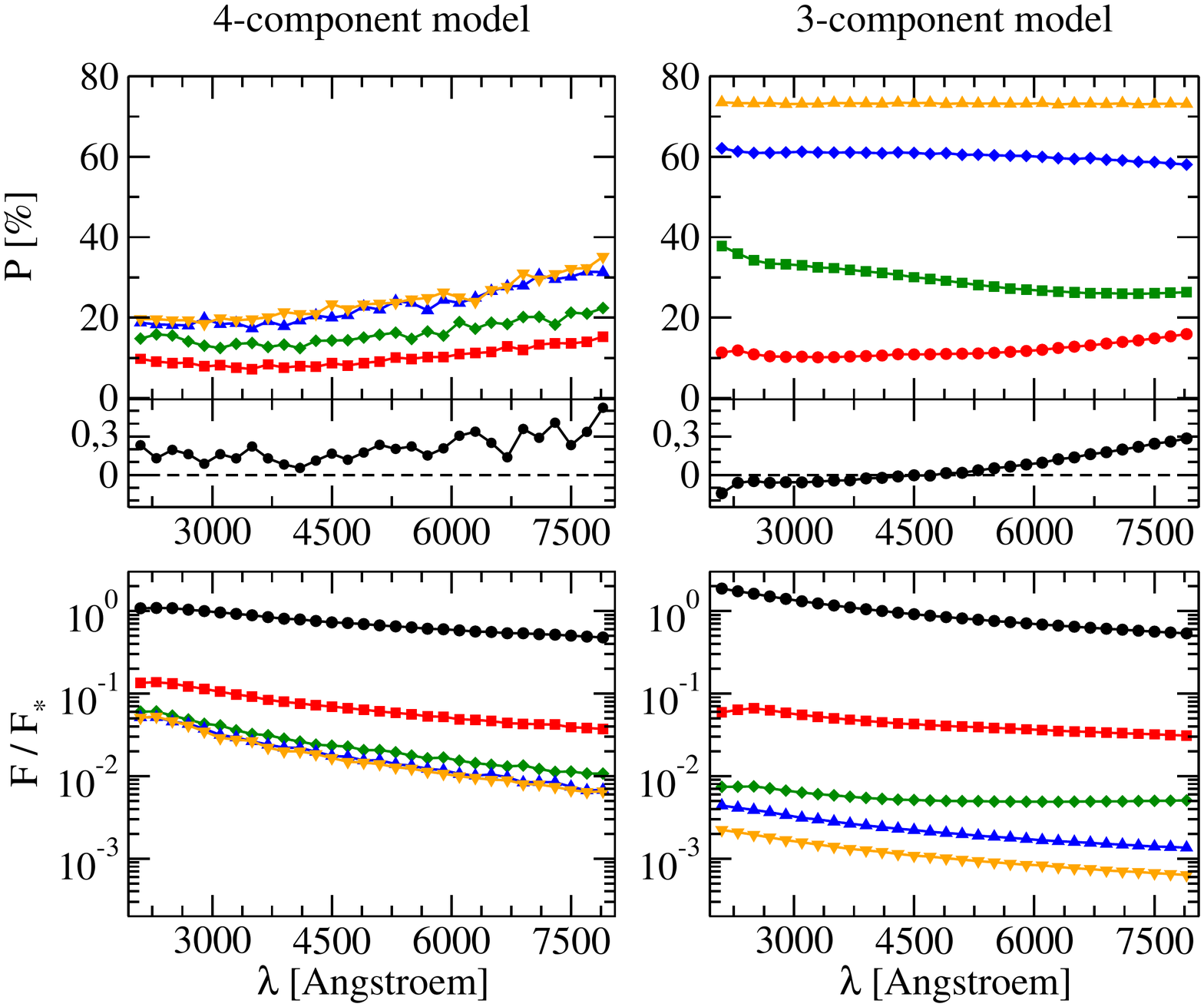} 
   \caption{Modeling the unified scheme of a thermal AGN.                               
	\textit{Left column} : the 4-component model as described in the text. 
        \textit{Right column} : the 3-component model lacking the NLR
        region as presented in \citet{Marin2012}. Both figures show the
        polarization percentage, \textit{P}, as a function of viewing
        inclination \textit{i} measured from the torus axis and 
        the fraction, $F/F_{\rm *}$, of the central flux, $F_{\rm *}$; Legend:
        $i$~=~$84^\circ$ (orange triangles with points to the bottom),
        $i$~=~$73^\circ$ (blue triangles with points to the top),
        $i$~=~$60^\circ$ (green diamonds), $i$~=~$46^\circ$ (red squares),
        $i$~=~$26^\circ$ (black circles).}
   \label{Fig1}%
\end{figure}

The resulting polarization percentage $P$ and the fraction, $F/F_{\rm *}$, of the central flux, $F_{\rm *}$, 
as a function of the viewing angle for both the 3-component and the 4-component model are shown in Fig.\ref{Fig1}. 
As expected, the overall polarization behavior for the model including the NLR (Fig.\ref{Fig1}, top left) is 
different from the results obtained for the 3-component model (Fig.\ref{Fig1}, top right). When adding the NLR, 
the global amount of polarization in type-2 objects decreases due to a combination of effects; the polarized flux 
from the ionized outflows is diluted by the relatively weaker polarization of the flux coming from the NLR. Note 
that for our dust mixture \citep{Mathis1977} the theoretical polarization degree produced by a single Mie scattering 
event reaches a maximum value of 35\% at 8000 \AA, while for electron scattering $P$ is as high as 100\%. The 
dilution from polar Mie scattering in the NLR is particularly efficient at edge-on viewing angles and $P$ decreases 
by a factor of $\sim 3$ in comparison to the 3-component model. Additional depolarization occurs because the photons 
passing through the NLR have to undergo more scattering events before they escape from the model region. Due to the 
wavelength-dependence of Mie scattering, the presence of the NLR also has an effect on the spectral shape of $P$.

Compared to the observed optical/UV polarization of Seyfert-2 galaxies \citep[][$P_{\rm obs} \le 10\%$]{Kay1994}, 
our theoretical polarization is still too high, even though the presence of the NLR helps to approach the observed 
range of $P$. As discussed in \citet{Marin2012}, our model so far assumes a uniform density in all reprocessing 
regions. However, there are observational hints that the torus and the polar outflows should be fragmented 
\citep{Nenkova2008,Nenkova2010}. Preliminary modeling of such a clumpy reprocessing structure with {\it STOKES} 
indicates that the resulting polarization further decreases with respect to a uniform density model and may allow 
us to explain the observations of \citet{Kay1994}. Apart from the polarization, the bottom of Fig.~\ref{Fig1} 
shows that the total (polarized + non-polarized) spectrum $F/F_{\rm *}$ decreases towards longer wavelengths in 
both models. This is another signature of the wavelength-dependent Mie scattering cross section and phase function. 
Also, a faint scattering feature at 2175~\AA~due to the carbonacious dust component is visible. When the NLR is 
included, the additional polar dust scattering increases the total flux scattered towards edge-on inclinations.

For type-1 objects our modeling shows that the additional dust scattering in the NLR can switch the $\vec E$-vector 
from a parallel to a perpendicular polarization state. While for the 3-component model the polar view exhibits parallel 
polarization below 4500~\AA~(Fig.\ref{Fig1}, top right, lower panel), the polarization is perpendicular across the
whole wavelength range in the 4-component case (Fig.\ref{Fig1}, top left, lower panel). Given the resulting weak 
percentage (< 0.5\%) of the perpendicular polarization, this model setup is coherent with the so-called 
\textit{polar scattering dominated} AGN \citep{Smith2002}. Such type-1 AGN present a perpendicular, optical polarization 
rising towards the blue that \citet{Smith2002} explained by dust extinction along a line of sight passing through the 
upper layers of the torus material. Our model qualitatively produces a similar wavelength-dependent polarization, but 
at a much lower viewing angle. However, the relatively low optical depth of the polar dust scattering does not reproduce 
the full extend of the observed rise in $P$ towards the blue end of the spectrum.

\section{Summary and conclusions}

To study the impact of the NLR on the polarization spectra in the optical/UV, we presented results for a 4-component 
reprocessing model computed with {\it STOKES}. In comparison with the 3-component model presented in \citet{Marin2012}, 
where polar dust scattering was absent, it appeared that the NLR cloud globally decreases the amount of polarization at 
type-2 viewing angles and may enforce perpendicular polarization also at type-1 viewing angles. The spectral shape of 
the polarization changes due to the additional Mie scattering in the polar directions.

An important conclusion to draw from this initial investigation is that the production of parallel polarization at 
type-1 viewing angles may become more difficult when the NLR is taken into account. It remains to investigate in a 
forthcoming study, to which extend the scattering origin of parallel polarization in type-1 AGN can be maintained.
Inclusion of clumpy reprocessing media is important in such an investigation. On the positive side, the NLR helps to 
explain the observed degree of continuum polarization in type-2 objects.

The NLR region is an important ingredient of the unified AGN scheme, and if the extended outflows in Seyfert-2 galaxies 
are indeed photo-ionization by the emission of an obscured primary source, then the NLR is expected to sustain the same 
half-opening angle as the obscuring circumnuclear matter \citep{Bianchi2012}. Constraints on the half-opening angle of 
the NLR could thus provide estimates of the relative ratio of Seyfert-1 to Seyfert-2 galaxies in the local universe. 
The geometrical shape of the NLR region may also help us to determine if the symmetry implied by the unified model is 
broken \citep[see][for a study on NGC~1068]{Raban2009,Goosmann2011,Marin2012a}. Such an asymmetry may explain the high 
degree of parallel polarization detected in peculiar objects like Mrk~231 \citep{Gallagher2005} and ultimately lead to 
a better understanding of the launching of outflows inside the funnel of the circumnuclear medium.

\acknowledgements
This research was supported by the French {\it GdR} PCHE and the research grant ANR-11-JS56-013-01.

\bibliographystyle{aa} 
\bibliography{marin} 

\end{document}